\documentstyle[12pt]{article}
\begin{document}

\newcommand\beq{\begin{equation}}
\newcommand\eeq{\end{equation}}
\newcommand\bea{\begin{eqnarray}}
\newcommand\eea{\end{eqnarray}}
\newcommand\df{$\delta$-function }
\newcommand\wf{wave functions }
\newcommand\pet{perturbation theory }

\begin{center}
{\bf Perturbation Theory for Singular Potentials in Quantum Mechanics}
\end{center}

\vskip 1 true cm
\centerline{\bf Diptiman Sen \footnote{E-mail address: 
diptiman@cts.iisc.ernet.in}} 

\vskip .5 true cm
\centerline{\it Centre for Theoretical Studies,}  
\centerline{\it Indian Institute of Science, Bangalore 560012, India} 

\vskip 1 true cm
\begin{abstract}

We study perturbation theory in certain quantum mechanics problems in
which the perturbing potential diverges at some points, even though
the energy eigenvalues are smooth functions of the coefficient of
the potential. We discuss some of the unusual techniques which are required 
to obtain perturbative expansions of the energies in such cases. These
include a point-splitting prescription for expansions around the
Dirichlet (fermionic) limit of the $\delta$-function potential, and
performing a similarity transformation to a non-Hermitian potential
in the Calogero-Sutherland model. As an application of the first technique,
we study the ground state of the $\delta$-function Bose gas near the 
fermionic limit.

\end{abstract}


\newpage

\section{Introduction}

Perturbation theory is an old and often used method for studying 
problems in quantum mechanics which are close to some exactly solvable
problems \cite{baym}. However, there are many models in which 
the traditional formulas of \pet lead to divergent 
expressions for the energies even at fairly low orders, although the actual
energy eigenvalues are known to be perfectly finite numbers. This situation 
typically (but not always) arises if the perturbing potential is singular, that
is, if it blows up at certain points in the configuration space. We are
therefore forced to develop new kinds of perturbative techniques if
we wish to expand such problems around the exactly solvable points.

We will discuss two such techniques here. The first one is used in the
presence of a \df potential in one dimension \cite{anez}, of which the \df 
Bose gas is a more general case \cite{lieb}. 
We will find that an expansion around the Neumann (bosonic) point, where the
coefficient of the \df potential is zero, is rather straightforward.
However, an expansion around the Dirichlet (fermionic) point, where the
coefficient is infinity, diverges at second order in perturbation theory.
We can obtain finite answers at that order by using a point-splitting
prescription. After discussing all this in a one-particle model in Secs. 2
and 3, we generalise the discussion to the \df Bose gas in Sec. 4. 

The other technique discussed is used in the case where the perturbing 
potential has a scale-invariant inverse-square form; this leads to
the \wf vanishing as a power of the two-particle separation whenever a pair
of particles approach each other. An important example is the 
Calogero-Sutherland model in one dimension \cite{calo,suth}, which has been 
studied extensively in recent years from many points of view (see \cite{sen1}
for a partial list of references). In Sec. 5, we show 
that this model can be studied perturbatively around the noninteracting
bosonic limit by performing a similarity transformation of the Hamiltonian
which takes into account the power-law form of the \wf near coincident points. 
This leads to a perturbation which has a non-Hermitian form; 
one can then employ the usual Rayleigh-Schr$\ddot o$dinger
perturbation theory. As a byproduct, we prove that the energies get no 
contributions beyond the leading one or two orders in \pet when the 
external (one-body) potential is either simple harmonic or zero; the \wf 
however do get contributions to higher orders. 

Historically, the method of using a non-Hermitian perturbation was perhaps 
first used to study models of anyons in two dimensions \cite{lein}, in order 
to study the spectrum near the bosonic limit \cite{sen2}. There is a vast
literature on this subject, so we will not discuss it here. We should
also briefly mention that there are other treatments of perturbation theory 
for singular potentials in two and higher dimensions, either in quantum 
mechanics \cite{manu} or in Chern-Simons field theories \cite{berg}. We will 
only consider one-dimensional models here which are technically simpler 
but have not been discussed as extensively.

In this work, we will not be exhaustive in our list of examples of singular 
potentials. Further, certain parts of our discussion are not new, and all
the models we will study are exactly solvable. We should therefore explain the
motivations behind our presentation. Our main aim is to introduce the reader 
to a class of problems in
which the usual form of \pet fails, and to indicate some
ways of dealing with such problems. A more complete study of this subject 
would certainly be worthwhile but we will not attempt that here.
Secondly, the models considered here would
not remain exactly solvable if we introduced additional non-singular
interactions or external potentials; however, such additional terms would
not change the nature of the new perturbative techniques which are designed
to address the singular parts of the interactions. It is therefore better to
study the nature of these new techniques in exactly solvable models so
that we can easily check that these methods do give the right answers.

In our discussion, we will use the following expression from 
Rayleigh-Schr$\ddot o$dinger \pet to quadratic 
order in a perturbation $V$. If $H=H_0 + V$, and the exact normalised 
eigenstates $\psi_n (0)$ and eigenvalues $E_n (0)$ of $H_0$ are known,
then the eigenstates of $H$ to first order in $V$ and the eigenvalues to 
second order in $V$ are given by \cite{baym}
\bea
\psi_n ~&=&~ \psi_n (0) ~+~ \sum_{l \ne n} ~\psi_l (0) ~\frac{\langle \psi_l 
(0) \vert V \vert \psi_n (0) \rangle}{E_n (0) - E_l (0)} ~, \nonumber \\
E_n ~&=&~ E_n (0) ~+~ \langle \psi_n (0) \vert V \vert \psi_n (0) \rangle ~
+~ \sum_{l \ne n} ~\frac{\langle \psi_n (0) \vert V \vert \psi_l (0) \rangle 
\langle \psi_l (0) \vert V \vert \psi_n (0) \rangle}{E_n (0) - E_l (0)} ~.
\label{pt}
\eea
If there is a degeneracy at energy $E_n (0)$, then we have to modify the
procedure at first order in $V$ by first diagonalising the matrix 
$V_{ln} = \langle \psi_l (0) \vert V \vert \psi_n (0) \rangle$. However, 
we will not have to face this problem in any of the models discussed in 
this paper, because the matrix elements $V_{ln}$ will always turn out to 
be zero for any two states for which $E_n (0) = E_l (0)$.

In Sec. 5, we will also need the general form of \pet at second and higher 
orders. The only information we will actually use is that, at all orders
in $V$ higher than the first, the change in the energy $E_n$ of the state 
$\psi_n$ involves products of matrix elements like
\beq
\langle \psi_n (0) \vert V \vert \psi_{l_k} (0) \rangle ~
\langle \psi_{l_k} (0) \vert V \vert \psi_{l_{k-1}} (0) \rangle ~...
\langle \psi_{l_2} (0) \vert V \vert \psi_{l_1} (0) \rangle ~
\langle \psi_{l_1} (0) \vert V \vert \psi_n (0) \rangle ~,
\label{hopt}
\eeq
where $k \ge 1$, and all the intermediate states $l_1 , l_2 , ... , l_k$ 
are different from $n$ \cite{baym}.

\section{The $\delta$-function potential in a box}

We begin with a particle which is restricted to move in one dimension in 
the range $-1 \le x \le 1$.  At the two end points $\vert x \vert = 1$, we
demand that the \wf should vanish. We impose a \df potential at
the origin. The Hamiltonian is
\beq
H ~=~ - ~\frac{1}{2m} ~\frac{\partial^2}{\partial x^2} ~+~ \frac{c}{m} ~
\delta (x) ~,
\label{ham1}
\eeq
where $c$ must be a real constant for $H$ to be Hermitian. (We set Planck's 
constant $\hbar =1$). We will only consider the subspace of \wf which 
are even functions of $x$, since the
odd functions vanish at the origin and are therefore unaffected by the
\df potential. The \df is equivalent to imposing the boundary condition 
\beq
\psi^\prime (0+) ~=~ - ~\psi^\prime (0-) ~=~ c \psi (0) ~,
\label{bc}
\eeq
where the primes denote derivatives with respect to $x$, and $0+$ and $0-$
denote the limits $x$ approaching the origin from the right and left 
respectively. (Note that $c=0$ gives us the Neumann boundary condition
$\psi^\prime (0) =0$, while $c=\infty$ gives the Dirichlet boundary condition
$\psi (0) =0$). It is clear that this probem is exactly solvable; the 
(unnormalised) eigenstates and eigenvalues of $H$ are given by
\bea
\psi_n ~&=&~ \sin ~[~ \alpha_n (1- \vert x \vert) ~] ~,  \nonumber \\
E_n ~&=&~ \frac{\alpha_n^2}{2m} ~, \nonumber \\
{\rm where} \quad \alpha_n ~\cot \alpha_n ~&=&~ -~c ~.
\label{psien1}
\eea

The normalised eigenstates and energies for $c=0$ are given by
\bea
\psi_n (x;0) ~&=&~ \sin ~[~(n ~+~ \frac{1}{2}) \pi (1 ~-~ \vert x \vert)~] ~, 
\nonumber \\
E_n (0) ~&=&~ \frac{(n ~+~ \frac{1}{2})^2 \pi^2}{2m} ~, 
\label{psien2}
\eea
where $n=0,1,2,... $. If we expand the energies in (\ref{psien1}) around 
their values at $c=0$, the leading order terms are
\beq
E_n (c) ~=~ \frac{1}{2m} ~[~ (n+\frac{1}{2})^2 \pi^2 ~+~ 2c ~-~ \frac{c^2}{(n ~
+~ \frac{1}{2})^2 \pi^2} ~+~ ... ~] ~.
\label{en1}
\eeq

For $c= \infty$, the eigenstates and energies are
\bea
\psi_n (x; \infty) ~&=&~ \sin ~[~ (n ~+~ 1) \pi (1 ~-~ \vert x 
\vert) ~] ~,  \nonumber \\
E_n (\infty) ~&=&~ \frac{(n ~+~ 1)^2 \pi^2}{2m} ~, 
\label{psien3}
\eea
where $n=0,1,2,... $. The energies in (\ref{psien1}) can be expanded around 
their values at $c=\infty$ in powers of $1/c$ as
\beq
E_n (c) ~=~ \frac{(n+1)^2 \pi^2}{2m} ~[~ 1 ~-~ \frac{2}{c} ~+~ \frac{3}{c^2} ~
+~ ... ~] ~.
\label{en2}
\eeq

We will now study how the expansions in (\ref{en1}) and (\ref{en2}) can be
obtained by using \pet around $c=0$ and $\infty$ respectively. Using the 
expressions in (\ref{pt}) with $V=(c/m) \delta (x)$, we recover the
results in (\ref{en1}); at second order, we have to use the easily derivable 
identity
\beq
\sum_{l \ne n} ~\frac{4}{(n ~+~ \frac{1}{2})^2 ~-~ (l ~+~ \frac{1}{2})^2} ~
=~ - ~\frac{1}{(n ~+~ \frac{1}{2})^2} ~,
\eeq
where the sum over $l$ runs from $0$ to $\infty$. 

We now have to discover how to perform \pet around $c=\infty$ in powers of
$1/c$. We begin as follows. If $\psi_n (x; c)$ and $E_n (c)$ are the 
exact results for the problem at any value of $c$, then the Feynman-Hellmann 
theorem says that with the perturbation $V = (c/m) \delta (x)$,
\beq
\frac{dE_n}{dc} ~=~ \langle \psi_n \vert ~\frac{1}{m} ~\delta (x) ~\vert 
\psi_n \rangle ~=~ \frac{1}{m} ~\vert \psi_n (0; c) \vert^2 ~.
\label{fh}
\eeq
Using (\ref{bc}), this implies that 
\beq
\frac{dE_n}{d (1/c)} ~=~ - ~\frac{1}{m} ~\vert \psi_n^\prime (0+ ;c) 
\vert^2 ~.
\label{de}
\eeq
Therefore, the change in energy to first order in $1/c$ can be obtained by 
taking the perturbation to be 
\beq
{\tilde V} ~=~ - ~\frac{1}{2cm} ~\delta^{\prime \prime} (x) ~.
\label{pt2}
\eeq
Here, it is understood that when we compute $\langle \psi_n \vert {\tilde V}
\vert \psi_n \rangle$, i.e., when calculating $\psi^\prime (0; c)$, we must 
take the limit $x \rightarrow 0$ either always from the right ($0+$) or 
always from the left ($0-$). This is because $\psi (x; c)$ has a cusp at the 
origin for all $c >0$; hence 
the two limiting values of $\psi^\prime (0; c)$ are equal in magnitude but
have opposite signs.

Eq. (\ref{pt2}) is one of the main results of this paper. It has not been
discussed much in the literature before except for a brief mention in Ref.
\cite{chit} (there is a minor error in the relevant equation in that paper). 
It is straightforward to check that using this potential as the perturbation
does give the correct first order change in energy if we use the $c=\infty$
results in (\ref{psien3}) as the uperturbed wave functions. However, if use
(\ref{pt2}) in the second order expression for the energy in (\ref{pt}), we 
obtain a divergent sum. We therefore have to {\it regularise} the sum in some
way. We do this as follows. From the first order result for $\psi_n$ in 
(\ref{pt}), we find that for $x >0$,
\bea
\psi_n^\prime (x; c) ~=~ &-& ~ (n+1) \pi ~\cos ~[~(n+1) \pi (1 - x )~] 
\nonumber \\
&+& ~ \frac{2(n+1)\pi}{c} ~\sum_{l\ne n} ~\frac{(-1)^{l+n} (l+1)^2 ~\cos ~
[~(l+1) \pi (1- x )~]}{(n+1)^2 ~-~ (l+1)^2} ~.
\label{psi2}
\eea
It is clear that if we naively set $x=0$ in (\ref{psi2}), the sum over $l$
will diverge. 

We now use the result
\beq
\sum_{l=-\infty}^{\infty} ~\exp ~(ilx) ~=~ 2 \pi \sum_{p=-\infty}^{\infty} ~
\delta ~(x - 2 \pi p) ~,
\label{comp1}
\eeq
which follows from the completeness of the functions $\exp (ilx)$ in any
interval in $x$ of length $2\pi$.
The right hand side of (\ref{comp1}) is zero in the neighbourhood of $x=0$
except exactly at the point $x=0$. We therefore set the left hand side of
the equation equal to zero {\it in the limit} $x \rightarrow 0$, although 
it diverges precisely at $x=0$. This implies that 
\beq
\lim_{x \rightarrow 0} ~\sum_{l \ne n} ~\cos ~(l+1) x ~=~ -~ \frac{3}{2} ~,
\label{reg1}
\eeq
since $l$ runs from $0$ to $\infty$.
We now subtract (\ref{reg1}) from the summation in (\ref{psi2}), and then
take the limit $x \rightarrow 0+$, to obtain a finite result 
\bea
\psi_n^\prime (0+; c) ~&=&~ (-1)^n ~(n+1) \pi ~\Bigl[~ 1 ~-~ \frac{2}{c} ~
\Bigl( ~\frac{3}{2} ~ +~ (n+1)^2 ~\sum_{l \ne n} 
\frac{1}{(n+1)^2 ~-~ (l+1)^2} ~\Bigr) ~\Bigr] ~, \nonumber \\
&=&~ (-1)^n ~ (n+1) \pi ~[ ~1 ~-~ \frac{3}{2c} ~] ~.
\eea
Substituting this in (\ref{de}), we obtain the correct second order result
in (\ref{en2}). 

Basically, we see that our regularisation of the energies at second order 
consists of using 
the first order wave functions {\it not} at the point $x=0$, but rather
at the point $x=0+$ (or $0-$). In the language of quantum field theory,
this is called a point-splitting prescription. Note also that the second order 
change in the ground state energy ($n=0$) in (\ref{en2}) is {\it positive},
although the formula in (\ref{pt}) would suggest that it should be negative
(since $E_n (0) < E_l (0)$ for all $l \ne n$);
this peculiarity is a consequence of the regularisation. 

We will not pursue \pet beyond second order near $c=\infty$. It seems likely
that a similar point-splitting prescription will also work at higher orders,
but this requires a much more detailed analysis.

\section{Combination of $\delta$-function and simple harmonic potentialss}

We now consider another exactly solvable example of the regularised \pet 
described in the previous section. The motivation for studying a second
example is to confirm that the point-splitting prescription is 
sound, and that it also works if the particle feels a potential in
addition to the $\delta$-function. We take the additional potential to
be simple harmonic \cite{anez}. Thus
\beq
H ~=~ - ~\frac{1}{2m} ~\frac{\partial^2}{\partial x^2} ~+~ \frac{c}{m} ~\delta 
(x) ~+~ \frac{1}{2} ~m \omega^2 x^2 ~.
\label{ham2}
\eeq
Once again, we restrict our attention to the subspace of even wave functions.
The exact eigenstates and energies are given in Ref. \cite{anez}. In 
particular, the energies are given by the transcendental equation
\beq
\frac{\Gamma \Bigl( ~\frac{3}{4} ~-~ \frac{E}{2\omega} ~\Bigr)}{\Gamma 
\Bigl( ~\frac{1}{4} ~-~ \frac{E}{2 \omega} ~\Bigr)} ~=~ - ~\frac{c}{2} ~.
\label{en3}
\eeq

From (\ref{en3}), we can compute the first three terms in an expansion
for $E_n$ around $c=0$ and $c=\infty$. Near $c=0$, 
\beq
\frac{E_n}{\omega} ~=~ n ~+~ \frac{1}{2} ~+~ \frac{c}{\sqrt{\pi m \omega}} 
\frac{(n-1)!!}{n!!} ~-~ \frac{c^2}{2\pi m \omega} ~\Bigl( ~\frac{(n-1)!!}{
n!!} ~ \Bigr)^2 ~[~ \psi (\frac{n}{2} ~ +~ 1) ~-~\psi (\frac{n}{2} + \frac{
1}{2})~]~,
\label{en4}
\eeq
where $n=0,2,4,...$, we define
\bea
p ~!! ~& \equiv &~ p(p-2)(p-4)... 2 \quad {\rm if} \quad p \quad {\rm is ~~
even ~~ and} \quad \ge 2 ~, \nonumber \\
& \equiv &~ p(p-2)(p-4)... 1 \quad {\rm if} \quad p \quad {\rm is ~~
odd ~~ and} \quad \ge 1 ~, \nonumber \\
& \equiv &~ 1 \quad {\rm if} \quad p=0 \quad {\rm or} \quad -1 ~,
\eea
and $\psi (z) = d \log \Gamma (z) /dz$ is the digamma function \cite{abra} 
(not to be confused with the wave function $\psi_n$). Near $c=\infty$,
\beq
\frac{E_n}{\omega} ~=~ n ~+~ \frac{1}{2} ~-~ \frac{2}{c} \sqrt{\frac{m 
\omega}{\pi}} ~\frac{ n!!}{(n-1)!!} ~+~
\frac{2 m \omega}{c^2 \pi} ~\Bigl( ~\frac{n!!}{(n-1)!!} ~\Bigr)^2 ~[~ \psi 
(\frac{n}{2} ~ +~ 1) ~-~ \psi (\frac{n}{2} + \frac{1}{2}) ~]~ ,
\label{en5}
\eeq
where $n=1,3,5,...$. The difference of digamma functions in (\ref{en4}) and
(\ref{en5}) is given by 
\bea
\psi (\frac{n}{2} + 1) - \psi (\frac{n}{2} + \frac{1}{2}) ~&=&~ 2 ~\log 2 
\quad {\rm if} \quad n=0 ~, \nonumber \\
&=&~ 2 ~[~ \log 2 ~+~ \sum_{k=1}^n ~\frac{(-1)^k}{k} ~] \quad {\rm if}
\quad n \quad {\rm is ~~ even ~~ and} \quad \ge 2 ~, \nonumber \\
&=&~ 2 ~[~ -~ \log 2 ~-~ \sum_{k=1}^{n} ~\frac{(-1)^k}{k} ~] \quad {\rm if} 
\quad n \quad {\rm is ~~ odd ~~ and} \quad \ge 1 ~. \nonumber \\
& &
\label{dig}
\eea
We will now derive the expressions (\ref{en4}) and (\ref{en5}) using 
perturbation theory.

For $c=0$, the exact eigenstates and energies are
\bea
\psi_n ~&=&~ (m \omega)^{1/4} ~H_n (y) ~\exp ~(- y^2 /2) ~, \nonumber \\
E_n ~&=&~ (~ n ~+~ \frac{1}{2} ~) ~\omega ~,
\label{psien4}
\eea
where $n=0,2,4,...$, and $H_n$ is the Hermite polynomial of degree $n$ in the
dimensionless variable $y = x \sqrt{m \omega}$. To be explicit \cite{abra},
\beq
H_n (y) ~=~ \Bigl( ~\frac{n!}{2^n \sqrt{\pi}} ~\Bigr)^{1/2} ~\sum_{m=0}^{n/2}~
\frac{(-1)^m ~(2y)^{2m}}{(\frac{n}{2} -m)!~(2m)!} ~.
\label{hn1}
\eeq
We now use the second-order formula in (\ref{pt}) and discover that the 
results match (\ref{en4}) provided the following identities are true
\beq
\sum_{l \ne n} ~\frac{(l-1)!!}{l!!} ~\frac{1}{l-n} ~=~ \frac{1}{2} ~
\frac{(n-1)!!}{n!!} ~[~ \psi (\frac{n}{2} ~+~ 1) ~-~ \psi (\frac{n}{2} ~+~ 
\frac{1}{2}) ~] ~, 
\label{iden}
\eeq
where the sum over $l$ runs over $0,2,4,...$. For $n=0$, this identity is
given in Ref. \cite{grad}, but for $n \ge 2$, we have not yet found it in the
literature. However we have checked it numerically by computing the 
summation on the left hand side from $l=0$ to $L=2 \times 10^7$ and 
setting the remainder (from $l=L+2$ to $l=\infty$) equal to
$\sqrt{2/(\pi L)}$ by using Stirling's formula. We find that the two sides
of (\ref{iden}) match upto $10^{-9}$ for $n=0$ to $100$.

For $c=\infty$, we have the exact results
\bea
\psi_n ~&=&~ (m \omega)^{1/4} ~H_n (\vert y \vert) ~\exp ~(- y^2 /2) ~, 
\nonumber \\
E_n ~&=&~ (~ n ~+~ \frac{1}{2} ~) ~\omega ~,
\label{psien5}
\eea
where $n=1,3,5,...$, and $H_n$ is given by
\beq
H_n (y) ~=~ \Bigl( ~\frac{n!}{2^n \sqrt{\pi}} ~\Bigr)^{1/2} ~
\sum_{m=0}^{(n-1)/2}~
\frac{(-1)^m ~(2\vert y \vert)^{2m+1}}{(\frac{n-1}{2} -m)!~(2m+1)!} ~.
\label{hn2}
\eeq
On using \pet with (\ref{pt2}), we recover the first order ($1/c$) result in
(\ref{en5}), but we again obtain a divergence at second order. We therefore
use the point-splitting prescription and Eqs. (\ref{pt}) and (\ref{de}) 
to obtain the second order energy
\beq
\frac{E_n^{(2)}}{\omega} ~=~ \lim_{y \rightarrow 0+} ~\frac{16 m \omega}{c^2} ~
\frac{n!}{( (\frac{n-1}{2})!)^2 \pi^{3/4} 2^n} ~\sum_{l \ne n} ~\frac{\sqrt{
l!} ~ H_l^\prime (y)}{(\frac{l-1}{2})! ~2^{l/2} ~(n-l)} ~.
\label{en6}
\eeq
where we sum over $l=1,3,5,...$.
As before, we can extract a finite value for the limit $x \rightarrow 0+$ by
using the completeness relation for Hermite polynomials
\beq
\sum_{p=0}^{\infty} ~H_p^\star (z) ~H_p (y) ~\exp ~[~- \frac{1}{2} ~(~
y^2 ~+~ z^2 ~) ~] ~=~ \delta ~(y-z) ~.
\label{comp2}
\eeq
Setting $z=0$, we get
\beq
\lim_{y \rightarrow 0+} ~\sum_{p=0,2,4,...} ~\frac{\sqrt{p!}}{(
\frac{p}{2})! ~\pi^{1/4} 2^{p/2}} ~ H_p (y) ~=~ 0 ~.
\label{comp3}
\eeq
We use the differential relation $H_l^\prime (y) = \sqrt{2l} H_{l-1} (y)$ 
\cite{abra}, and subtract (\ref{comp3}) from (\ref{en6}). We then discover 
that (\ref{en6}) agrees with (\ref{en5}) provided again that the identities
in (\ref{iden}) hold.

The results in this section indicate the general method of regularising second
order \pet near $c=\infty$. Namely, we have to use the completeness relations
which hold for the solutions at $c=\infty$, as exemplified by the relations
(\ref{comp1}) and (\ref{comp2}), in order to obtain a series whose sum is 
actually finite in the limit $x \rightarrow 0+$, although it diverges 
if we naively set $x=0$. We then have to subtract this sum from the second 
order perturbation result in order to get a finite result in the limit $x 
\rightarrow 0+$.

\section{The $\delta$-function Bose gas}

Having convinced ourselves that \pet does work in the one-particle problem with
a \df potential, let us consider a many-body generalisation. We will study
the \df Bose gas which has been studied extensively since it was first solved
in Ref. \cite{lieb} using the Bethe ansatz. Although the exact results are 
available, it is instructive to see how they can be recovered by using \pet
near $c=0$ and $\infty$. We will consider $N$ particles on a circle of
circumference $L$. We will sometimes be interested in the thermodynamic limit
$N, L \rightarrow \infty$, keeping the density $\rho = N/L$ fixed. The 
Hamiltonian is
\beq
H ~=~ -~\frac{1}{2m} ~\sum_{1 \le i \le N} ~\frac{\partial^2}{\partial x_i^2}~
+~ \frac{2c}{m} ~\sum_{1 \le i < j \le N} ~\delta ~(~x_i ~-~ x_j ~) ~.
\label{ham3}
\eeq
Since the particles are identical bosons, the \wf must be completely 
symmetric. There are $N!$ possible orderings of the particle coordinates,
given by $0 \le x_{P_1} \le x_{P_2} \le ... \le x_{P_N} \le L$, where 
$(P_1,P_2,...,P_N)$ is some permutation of the numbers $(1,2,...,N)$. If
the \wf are known for any one ordering, say $0 \le x_1 \le x_2 \le ... x_N 
\le L$, they are known for all other orderings by symmetry.

Now it is clear that the model describes noninteracting bosons for $c=0$.
For $c=\infty$, the \wf vanish whenever any two particle coordinates coincide.
We can then carry out the unitary transformation $\psi_P \rightarrow (-1)^P 
\psi_P$,
where $\psi_P$ denotes the wave function for the ordering $P$, and $(-1)^P$
denotes the sign of the permutation $P$. Under this transformation, the
wave function becomes completely antisymmetric, i.e., fermionic. Thus 
$c=\infty$ denotes a system of noninteracting fermions. Note that this
unitary transformation is only allowed if $c=\infty$. At any other value of
$c$, the symmetric \wf do not vanish for $x_i = x_j$, and the transformation
would produce antisymmetric \wf which are discontinuous at those coincident
points.

For simplicity, we will limit our discussion only to the ground state wave 
function and the corresponding energy per particle. At $c=0$, these are 
given by
\bea
\psi_0 ~&=&~ \frac{1}{\sqrt{L^N}} ~, \nonumber \\
\frac{E_0}{N} ~&=&~ 0 ~.
\eea
We now consider the perturbative expansion around $c=0$. At first 
order, the change in energy is found to be
\beq
\frac{E_0^{(1)}}{N} ~=~ \frac{c ~(N-1)}{L} 
\eeq
which is equal to $c \rho$ in the thermodynamic limit. This agrees with 
the result in Ref. \cite{lieb}. At second order, there are non-zero matrix 
elements between $\psi_0$ and 
\beq
\psi_{k,-k} ~\equiv ~\frac{1}{\sqrt{L^N N(N-1)}} ~\sum_{i \ne j} ~\exp ~[~ik 
(x_i ~ -~ x_j)~] ~,
\eeq
where $k=2\pi n/L$ and $n=1,2,3,...$. The second order formula in (\ref{pt})
then gives 
\beq
\frac{E_0^{(2)}}{N} ~=~ -~ \frac{c^2 (N-1)}{6} ~.
\label{enx}
\eeq
This is indeed the correct answer (i.e., it agrees with the Bethe 
ansatz results) if we take the limit $c \rightarrow 0$ holding $N$ 
and $L$ {\it fixed}. However, it is {\it not} the correct result if we
hold $c$ fixed and take the thermodynamic limit, as can be seen from the fact
that the right hand side of (\ref{enx}) diverges in that limit. The reason 
for this divergence is that the matrix element 
\beq
\langle \psi_{k,-k} \vert V \vert \psi_0 \rangle ~=~ 
\frac{2c \sqrt{N(N-1)}}{L}
\eeq
can become comparable to or even larger than the energy denominator $E_0 -
E_{k,-k}$ if we let $L \rightarrow \infty$. Second order \pet breaks down
in that case. In the thermodynamic limit, a different \pet exists due to
Bogoliubov \cite{lieb}, and it gives the result
\beq
\frac{E_0^{(2)}}{N} ~=~ -~ \frac{4}{3 \pi} ~c^{3/2} ~\rho^{1/2} ~.
\eeq
The presence of a fractional power of $c$ clearly shows that this \pet 
is rather different from the one we have been discussing so far. We emphasize
that the divergence in our second order \pet in the thermodynamic limit
is not to be confused with the divergence which we have discussed in the
Secs. 2 and 3; the two divergences are due to entirely different 
reasons.

We will now consider \pet around the free fermion point $c=\infty$. Here we
will again discover a divergence at second order \pet which is similar
to the ones discussed in the previous two sections, and the regularisation 
required to cure it is also very similar. However, unlike the expansion around
$c=0$, there is no additional divergence in the thermodynamic limit, and
$E_0/N$ does have an expansion in integer powers of the dimensionless 
parameter $\rho /c$. 

At $c=\infty$, the ground state wave function $\psi_0$ in a particular 
ordering of the particle coordinates is given by $1/\sqrt{L^N N!}$ times the 
determinant of the matrix 
\beq
M_{ij} ~=~ \exp ~[~i k_i x_j ~] ~,
\eeq
where the momenta $k_i$ run over the $N$ distinct values $(\pi /L)~(1-N,
3-N,5-N,...,N-1)$. (Note that for $c=\infty$, the allowed values of the
momenta are given by $n 2\pi /L$ if $N$ is odd, and $(n+ \frac{1}{2}) 2\pi /L$
if $N$ is even \cite{lieb}). The ground state energy is given by
\beq
\frac{E_0}{N} ~=~ \frac{1}{2mN} ~\sum_{i} ~k_i^2 ~=~ \frac{\pi^2}{6mL^2} ~(~ 
N^2 ~-~ 1~) ~.
\label{en7}
\eeq
Following the arguments leading upto (\ref{pt2}), we can show that the 
perturbation around $c=\infty$ is described by the two-particle interaction
\beq
{\tilde V} ~=~ -~ \frac{1}{cm} ~\sum_{i<j} ~\delta^{\prime \prime} ( ~x_i ~
-~ x_j ~)~.
\label{pt3}
\eeq
(The simplest way to derive this is to consider just two particles, and 
transform to the centre of mass and relative coordinates $X=(x_1 + x_2)/2$ 
and $x=x_1 -x_2$ respectively. The contact interaction then involves only 
the coordinate $x$ which is a one-body problem as in the previous sections).
Using this potential in first order \pet yields the change in energy
\beq
\frac{E_0^{(1)}}{N} ~=~ -~ \frac{\pi^2}{3mcL^3} ~N~(~ N^2 ~-~ 1~) ~.
\label{en8}
\eeq

We now proceed to second order perturbation theory. To simplify the 
calculation, we will consider the thermodynamic limit; thus we will replace 
$\sum_{k_i}$ by $(L/2\pi) ~\int dk$ whenever convenient. It is useful 
to introduce the Fermi momentum
\beq
k_F ~=~ \pi \rho ~,
\label{fm}
\eeq
so that all the momenta $k_i$ in the ground state lie within the range 
$[-k_F , k_F]$.
We begin our analysis by observing that $\tilde V$ has non-zero matrix 
elements between the ground state $\psi_0$ and the excited states 
$\psi_{k_i^\prime , k_j^\prime}$,
where $\psi_{k_i^\prime , k_j^\prime}$ differs from $\psi_0$ in having exactly
two momenta $k_i^\prime$ and $k_j^\prime$ different from two momenta $k_i$
and $k_j$; the other $N-2$ momenta are identical in the two states. The 
translation invariance of the interaction (\ref{pt3}) implies that we 
must have 
\beq
k_i ~+~ k_j ~=~ k_i^\prime ~+~ k_j^\prime ~,
\label{mc}
\eeq
where $k_i^\prime$ and $k_j^\prime$ must lie {\it outside} the range 
$[-k_F , k_F]$.

We then find that the second order energy is given by
\beq
\frac{E_0^{(2)}}{N} ~=~\frac{2}{mc^2 NL^2} ~\sum_{k_i < k_j} ~(~ k_i ~-~ k_j ~
)^2 ~\sum_{k_i^\prime < k_j^\prime} ~\frac{(~ k_i^\prime ~-~ k_j^\prime ~)^2}{
k_i^2 ~+~ k_j^2 ~-~ k_i^{\prime 2} ~-~ k_j^{\prime 2}} ~,
\label{en9}
\eeq
which is again divergent since the sum over $k_i^\prime$ or $k_j^\prime$ runs 
over all values outside the range $[-k_F , k_F]$. Note that there are only 
three independent sums on the right hand side of (\ref{en9}) since there is a
constraint from momentum conservation (\ref{mc}); thus we can replace the
sums by multiple integrals over $k_i$, $k_j$ and $k_j^\prime$ of the form
\beq
\int_{-k_F}^{k_F} ~dk_i ~\int_{k_i}^{k_F} ~dk_j ~\int_{k_0}^{\infty} ~
dk_j^\prime ~,
\label{int1}
\eeq
where the lower limit of the $k_j^\prime$ integral is given by
\bea
k_0 ~&=&~ k_F \quad {\rm if} \quad k_i ~+~ k_j ~<~ 0 ~, \nonumber \\
&=&~ k_F ~+~ k_i ~+~ k_j \quad {\rm if} \quad k_i ~+~ k_j ~>~ 0~.
\label{lim}
\eea

We regularize the divergence in (\ref{en9}) using the method described 
in Sec. 2. We will
briefly indicate the procedure here without going through all the details.
In the matrix element $\langle \psi_0 \vert {\tilde V} 
\vert \psi_{k_i^\prime , k_j^\prime} \rangle$, let us consider one of the
terms in ${\tilde V}$, say, $\delta^{\prime \prime} (x_1 - x_2)$. After
performing the integrals over $x_3 , x_4 , ... , x_N$, we are left with 
integrals over $X = (x_1 + x_2 )/2$ and $x=x_1 - x_2$ of the form $\int_0^L
dX/L$ times
\beq
\int dx ~\delta (x) ~\cos ~[\frac{1}{2} (k_i^\prime - k_j^\prime ) x] ~\cos ~
[\frac{1}{2} (k_i - k_j ) x] ~=~ \frac{1}{2} \int dx ~\delta (x) ~[ \cos ~
(k_j^\prime -k_i )x ~+~ \cos ~(k_j^\prime - k_j )x ]
\label{int2}
\eeq
appearing in the numerator of the last term on the right hand side of
(\ref{en9}). If we naively perform the integral in 
(\ref{int2}) by setting $x=0$, we obtain the constant 
$1$; this leads to the divergence in (\ref{en9}) when we integrate a constant
over $k_j^\prime$ with the ranges shown in (\ref{int1}) and (\ref{lim}). To 
regularise that divergence, we first perform the integral over $k_j^\prime$ 
for $x \ne 0$, and then take the limit $x \rightarrow 0+$. We use the result 
\beq
\frac{1}{2} ~\int_{k_0}^{\infty} ~dk_j^\prime ~[~ \cos ~(k_j^\prime -k_i )x ~
+~ \cos ~( k_j^\prime - k_j )x ~] ~=~ - ~k_F ~-~ \frac{1}{2} ~\vert ~k_i ~+~ 
k_j ~\vert 
\label{int3}
\eeq
in the limit $x \ne 0$. This follows from the fact that 
\beq
\int_{-\infty}^{\infty} ~dk ~\exp ~[ikx] ~=~ 2 ~\int_{0}^{\infty} ~dk ~
\cos ~(kx) ~=~ 0 
\eeq
if $x \ne 0$, and therefore
\beq
\int_{K}^{\infty} ~dk ~\cos ~(kx) ~=~ -~ \frac{\sin ~(K x)}{x} 
\eeq
which equals $-K$ in the limit $x \rightarrow 0+$.

After subtracting the integral in (\ref{int3}) from the right hand
side of (\ref{en9}), we are left with the convergent expression
\bea
\frac{E_0^{(2)}}{N} ~&=&~ \frac{1}{4 \pi^3 mc^2 \rho} ~\int_{-k_F}^{k_F} ~
dk_i ~\int_{k_i}^{k_F} ~dk_j ~(~k_i ~-~k_j ~)^2 ~I (k_1 , k_2 ) ~,
\nonumber \\
{\rm where} \quad I (k_1 , k_2 ) ~&=& ~2k_F ~+~ \vert ~ k_i ~+~ k_j ~
\vert ~-~ \frac{1}{2} ~(k_i - k_j )^2 ~ \int_{k_0}^{\infty} ~
dk_j^\prime ~\frac{1}{(k_j^\prime - k_i)(k_j^\prime - k_j)} ~. \nonumber \\
& &
\label{int4}
\eea
On doing the integrals in (\ref{int4}) and using (\ref{fm}), we get
\beq
\frac{E_0^{(2)}}{N} ~=~ \frac{\pi^2 \rho^4}{2 m c^2} ~.
\eeq
Collecting all the results near $c=\infty$, we find the first three
terms in the energy per particle in the thermodynamic limit to be
\beq
\frac{E_0}{N} ~=~ \frac{\pi^2 \rho^2}{6m} ~[~ 1 ~-~ 2~\frac{\rho}{c} ~+~ 3~
\frac{\rho^2}{c^2} ~+~ ... ~] ~.
\eeq
This agrees with the results in \cite{lieb}.

\section{The Calogero-Sutherland model}

We now turn to another technique for dealing with singular potentials.
Consider the Calogero-Sutherland model in which particles move on a line
(or circle), and interact pairwise through an inverse-square potential.
This potential is so singular that the \wf vanish 
whenever two particles $i$ and $j$ approach each other. This
property can create difficulties for the usual kind of \pet if we perturb
around a model in which the inverse-square potential is absent and the
\wf do not vanish at coincident points. We will discuss two different 
versions of the inverse-square model, one with particles 
placed in a simple harmonic potential on a line, and the other with particles 
on a circle with no external potential. The modification of \pet required
in the two cases is similar, but it is convenient to discuss them separately.

\subsection{The Calogero model}

Consider the Hamiltonian for $N$ identical particles on a line \cite{calo}
\beq
H ~=~ -~\frac{1}{2m} ~\sum_i \frac{\partial^2}{\partial x_i^2} ~+~ \frac{
\lambda (\lambda -1)}{m} ~\sum_{i<j} ~\frac{1}{(x_i ~-~ x_j)^2} ~+~ \frac{
m \omega^2}{2} ~\sum_i ~x_i^2 ~,
\label{ham4}
\eeq
where $\lambda \ge 0$. To make the problem completely well-defined, we have
to specify that all \wf vanish as $\vert x_i - x_j \vert^{\lambda}$ whenever
particles $i$ and $j$ approach each other. Since the singular two-particle
interactions prevent particles from crossing each other, there is no way of 
comparing the phase of the wave function for one configuration with the phase
for another configuration in which two particles have been exchanged. In other
words, we have the freedom to work with any phase convention relating the 
different possible orderings of the particle coordinates. We will choose to
work with completely symmetric wave functions. Note that the inverse-square 
interaction in (\ref{ham4}) vanishes for both $\lambda =0$ and
$\lambda =1$. However, the condition on the \wf implies that $\lambda =0$ 
corresponds to noninteracting bosons, while $\lambda =1$ corresponds to
noninteracting fermions. The latter is true even though we are working with 
symmetric wave functions; the situation is very similar to the
model in Sec. 4 with $c=\infty$.

It is particularly simple to solve the model at the two noninteracting
values $\lambda =0$ and $1$, and we can consider doing \pet about either
point. It turns out that \pet around $\lambda =1$ is completely straightforward
since all matrix elements of the perturbation (taken to be the inverse-square
term in (\ref{ham4})) are finite. However, \pet around $\lambda =0$ is not so
simple because even the first order result in (\ref{pt}) diverges. This can
be seen from a simple counting of powers; if we define a relative coordinate
$x = x_1 - x_2$, the \wf for $\lambda =0$ go to a non-zero constant as
$x \rightarrow 0$. Then first order \pet involves an integral like
$\int dx /x^2$ which diverges at $x=0$. (The same power counting shows that
\pet around any value of $\lambda \le 1/2$ is divergent; the divergence
is weakest (logarithmic) at $\lambda =1/2$ \cite{amel}).

We therefore proceed as follows for \pet around $\lambda =0$. Let us denote the
singular part of the \wf at any value of $\lambda$ by
\beq
\Delta (\lambda) ~=~ \prod_{i<j} ~\vert ~x_i ~-~ x_j ~\vert^{\lambda} ~.
\eeq
Instead of solving $H \psi = E \psi$, we will solve ${\tilde H} {\tilde
\psi} = E {\tilde \psi}$, where ${\tilde \psi} = \Delta^{-1} \psi$ does 
{\it not} vanish at coincident points, and
\bea
{\tilde H} ~&=&~ \Delta^{-1} ~H~ \Delta ~=~ H_0 ~+~ {\tilde V} ~, \nonumber \\
{\rm where} \quad H_0 ~&=&~  -~\frac{1}{2m} ~\sum_i \frac{\partial^2}{\partial 
x_i^2} ~+~ \frac{ m \omega^2}{2} ~\sum_i ~x_i^2 ~, \nonumber \\
{\rm and} \quad {\tilde V} ~&=&~  -~ \frac{\lambda}{m} ~\sum_{i<j} ~
\frac{1}{x_i - x_j}~ (~\frac{\partial}{\partial x_i} ~-~ \frac{\partial}{
\partial x_j} ~) ~.  
\label{ham5}
\eea

We now have to derive the formulas for \pet with the {\it non-Hermitian}
interaction $\tilde V$ given in (\ref{ham5}). We can show that the required 
formulas are identical to those appearing in the usual 
Rayleigh-Schr$\ddot o$dinger perturbation theory, such as 
Eqs. (\ref{pt}) and (\ref{hopt}) \cite{baym}. We only have to remember to 
use the expressions exactly as given in (\ref{pt}) and schematically as given
in (\ref{hopt}). For instance, we cannot replace the matrix element 
$\langle \psi_n (0) \vert {\tilde V} \vert \psi_l (0) \rangle$ by the 
complex conjugate of $\langle \psi_l (0) \vert {\tilde V} \vert \psi_n (0) 
\rangle$ since $\tilde V$ is not Hermitian. This will become clearer 
after the discussion below.

The exact eigenstates and eigenvalues of the Hamiltonian $H_0$ have the general
form
\bea
\Psi_n (x_i ~;~ \lambda =0) ~&=&~ P_n (x_i) ~G ~, \nonumber \\
G ~&=&~ \exp ~[~-~ \frac{m\omega}{2} ~\sum_i ~ x_i^2 ~] ~, \nonumber \\
E_n^{(0)} ~&=&~ (~ n ~+~ \frac{N}{2} ~) ~\omega ~,
\eea
and $P_n$ is a symmetric polynomial in the $x_i$ of degree $n$. In order to 
discuss perturbation theory, it is convenient to use the following result
\cite{gura}. Any symmetric polynomial $Q_n (x_i)$ of degree $n$ can be 
written as a linear combination 
\beq 
Q_n (x_i) ~=~ \sum_l ~P_l (x_i) ~,
\label{super}
\eeq
where the symmetric polynomials $P_l$ all have degree $l \le n$ and the \wf
$P_l G$ are eigenstates of $H_0$.

Let us now consider first order perturbation theory. We note that
\beq
{\tilde V} ~(~ P_n ~G ~) ~=~ \frac{\lambda N (N-1)}{2} ~P_n ~G ~+~ Q_l ~G ~,
\label{vp}
\eeq
where $Q_l$ is a polynomial of degree $\le n-2$. Using (\ref{super}), $Q_l G$ 
is seen to be a superposition of eigenstates of $H_0$ with energy less than 
that of $P_n G$. Therefore $P_n G$ is orthogonal to $Q_l G$. Hence the 
first order change in energy is just
\beq
E_n^{(1)} ~=~ \frac{\lambda N (N-1)}{2} ~.
\label{en10}
\eeq
One might worry about degeneracies at this stage. To see that this is not a
problem, consider two polynomials $P_n^{(1)}$ and $P_n^{(2)}$ of the same 
degree $n$ such that $\psi^{(1)} = P_n^{(1)} G$ and $\psi^{(2)} = P_n^{(2)} 
G$ are degenerate and
orthogonal eigenstates of $H_0$. The orthogonality together with the form
in (\ref{vp}) shows that both the matrix elements $\langle \psi^{(1)} \vert 
{\tilde V} \vert \psi^{(2)} \rangle$ and $\langle \psi^{(2)} \vert {\tilde V} 
\vert \psi^{(1)} \rangle$ are zero. It is therefore unnecessary to worry about 
degenerate \pet at first order. We also have the general result that
$\langle \psi_l \vert {\tilde V} \vert \psi_n \rangle$ can be non-zero only if
either $\psi_l = \psi_n$ or if the degrees of the respective polynomials
satisfy $l \le n-2$. This result will be used below.

We now consider higher order perturbation theory. From the general form
given in (\ref{hopt}), it is easy to see that there are {\it no} contributions
to the energies at any order higher than the first. This is because 
the individual matrix elements in (\ref{hopt}) can only be non-zero
if the degrees of the polynomials satisfy $n \le l_k -2$, $l_k \le 
l_{k-1}$, ..., $l_2 \le l_1$ and $l_1 \le n-2$. (These inequalities follow 
from the observation that the states $\psi_{l_i}$ may be identical to each 
other, but they all have to be different from $\psi_n$). It is clear that
all these inequalities cannot hold simultaneously; hence the product in
(\ref{hopt}) must necessarily be zero.

Our \pet has therefore yielded the well-known result that the energies only
get a contribution at first order in $\lambda$, and that this contribution is
the same for all states as seen in (\ref{en10}). However the eigenstates
can get contributions at any order in $\lambda$.

Before ending this subsection, we would like to emphasise that although 
the perturbation $\tilde V$ is not Hermitian, its derivation from 
the Hamiltonian in (\ref{ham4}) guarantees that the perturbative changes in
energies must be real. Note also that the eigenstates of $\tilde H$ are 
orthogonal with the integration measure $\Delta^2 \prod_i dx_i$, not with 
the measure $\prod_i dx_i$. 

\subsection{The Sutherland model}

We now consider the inverse-square model on a circle of circumference $L$
with periodic boundary conditions \cite{suth}. The Hamiltonian for $N$ 
particles is
\beq
H ~=~ -~\frac{1}{2m} ~\sum_i ~\frac{\partial^2}{\partial x_i^2} ~+~ \frac{
\lambda (\lambda -1)}{m} ~\sum_{i<j} ~\frac{\pi^2}{L^2 \sin^2 ~[~\frac{\pi}{
L} ~(~x_i ~-~ x_j ~) ~]} ~.
\label{ham6}
\eeq
We impose the same conditions on the \wf at coincident points as in the 
previous subsection. To develop perturbation theory, we again perform
a similarity transformation but now with a periodic function
\beq
\Delta (\lambda) ~=~ \prod_{i<j} ~\vert \sin ~[~ \frac{\pi}{L} ~(x_i ~-~ x_j) ~
]~ \vert^{\lambda} ~.
\eeq
This transforms the Hamiltonian into
\bea
{\tilde H} ~&=&~ \Delta^{-1} ~H~ \Delta ~=~ H_0 ~+~ {\tilde V} ~, \nonumber \\
{\rm where} \quad H_0 ~&=&~  -~\frac{1}{2m} ~\sum_i \frac{\partial^2}{\partial 
x_i^2} ~+~ \frac{\lambda^2 \pi^2 N (N^2 -1)}{6mL^2} ~, \nonumber \\
{\rm and} \quad {\tilde V} ~&=&~  -~ \frac{\lambda}{m} ~\frac{\pi}{L} ~
\sum_{i<j} ~\cot ~[~ \frac{\pi}{L} (x_i - x_j)~]~ (~\frac{\partial}{\partial
x_i} ~-~ \frac{\partial}{\partial x_j} ~) ~.
\label{ham7}
\eea
Note that $H_0$ has a $\lambda$ dependent constant which will therefore
appear in the energies of all the states. We now discuss \pet with the 
interaction $\tilde V$. 

The (unnormalised) eigenstates and eigenvalues of the noninteracting 
Hamiltonian $H_0$ are easy to derive. Consider a set of $N$ integers $\{ n_i
\} = (n_1 , n_2 , ... , n_N)$, each of which may be positive, negative or 
zero, and some of them may be equal to each other. Then we have
\bea
\psi_{\{ n_i \}} ~&=&~ \exp ~[~i \frac{2\pi}{L} ~(n_1 x_1 + n_2 x_2 + ... + 
n_N x_N ~)~] ~+~ \quad {\rm symmetrisation} ~, \nonumber \\
E_{\{ n_i \}} ~&=&~ \frac{2 \pi^2}{mL^2} ~\sum_i ~n_i^2 ~.
\label{psien6}
\eea
The momentum of such a state is given by $(2\pi /L)\sum_i n_i$.

In order to discuss perturbation theory, it is convenient to introduce the
concept of {\it squeezing} \cite{suth}. We say that a set of integers 
$\{ n_i^\prime \}$ can be obtained from another set $\{ n_i \}$ by squeezing
if (a) two integers in the first set, say, $n_i^\prime$ and $n_j^\prime$,
are different from two integers in the second set, say, $n_i$ and $n_j$,
but all the other $N-2$ integers in the two sets are pairwise equal, and (b)
$n_i^\prime + n_j^\prime = n_i + n_j$ but $\vert n_i^\prime - n_j^\prime
\vert < \vert n_i - n_j \vert$. Thus the momentum of the states corresponding
to the two sets are equal, but $E_{\{ n_i^\prime \}} < E_{\{ n_i \}}$ 
according to (\ref{psien6}).

We then find that the action of $\tilde V$ on the states is given by
\beq
{\tilde V} ~\psi_{\{ n_i \}} ~=~ \Bigl( ~\frac{2 \lambda}{m} \frac{\pi^2}{
L^2} ~\sum_{i<j} ~\vert ~n_i ~-~ n_j ~\vert ~\Bigr) ~\psi_{\{ n_i \}} ~+~
\sum_{\{ n_i^\prime \}} ~\psi_{\{ n_i^\prime \}} ~,
\eeq
where the summation over states runs over the various possible sets $\{ 
n_i^\prime \}$ which can be obtained from the set $\{ n_i \}$ by squeezing;
all these states have lower energy than $\psi_{\{ n_i \}}$ by the comment
in the previous paragraph. We thus see that a matrix element $\langle
\psi_{\{ n_i^\prime \}} \vert {\tilde V} \vert \psi_{\{ n_i \}} \rangle$ can 
only be non-zero if $\{ n_i^\prime \}$ is either identical to $\{ n_i \}$  
or can be obtained from it by squeezing. Clearly, matrix elements between
distinct but degenerate states are zero, so that we do not need to do
any degenerate perturbation theory. 

The first order change in energy is therefore given by
\beq
E_{\{ n_i \}}^{(1)} ~=~ \frac{2 \lambda}{m} \frac{\pi^2}{L^2} ~\sum_{i<j} ~
\vert ~n_i ~-~ n_j ~\vert ~.
\label{en11}
\eeq
Once again, there is no contribution to the energies from any higher orders
in \pet due to the product form in (\ref{hopt}); at least one of the matrix 
elements in that product must be zero following an argument very similar
to the one used in the previous section (with the concept of "degree of
polynomial" replaced by "squeezing").

We have thus recovered the result that there are only two $\lambda$ dependent
terms in the energy of any state. The term of order $\lambda$ is given in 
(\ref{en11}) and it depends on the state, whereas the term of order 
$\lambda^2$ given in (\ref{ham7}) is the same for all states. 

\section{Outlook}

There are several issues which could be addressed in the future. For the \df 
problem near $c=\infty$ in Secs. 2-4, we could study \pet at cubic and higher
orders to see if a similar point-splitting regularisation can be devised to 
obtain finite answers at all orders in $1/c$. 

We can also study a generalisation of the \df problem to higher dimensions.
Given a bounded region $M$ with boundary $\partial M$, it is known \cite{chav}
that the Hamiltonian $H = - {\vec \nabla}^2$ is self-adjoint (and therefore
has real eigenvalues and orthonormal eigenvectors) if we impose the boundary 
condition on all wave functions,  
\beq
{\hat n} \cdot {\vec \nabla} \psi ~=~ c \psi
\eeq
where $\hat n$ is the unit vector pointing inward and normal to the boundary
$\partial M$, and $c$ can be any real number. Once again, we can
study \pet near $c=0$ (Neumann condition) or $c=\infty$ (Dirichlet
condition). To do this, we first define a generalised $\delta$-function
and its derivatives which are non-zero only in an infinitesimal 
neighbourhood of $\partial M$. We can then use expressions like (\ref{pt})
or (\ref{pt2}) where the perturbation $V$ or $\tilde V$ involves that 
$\delta$-function or its 
second derivative. Once again, we obtain a divergence in second order \pet 
near $c=\infty$, and a regularisation is required to obtain finite answers.
It would be interesting to study the precise form of the regularisation for an
arbitrary region $M$.

The method of non-Hermitian perturbation has already been used in 
two-dimensional models such as anyons \cite{sen2} and systems with other 
types of two-body correlations \cite{bhad}. We can consider applications
of this method to
any model in which conventional \pet diverges due to the singular nature of
the \wf in the vicinity of certain points in configuration space. We simply
remove the singular part of the \wf by a similarity transformation; we then
obtain a new potential (generally non-Hermitian) whose matrix elements 
between non-singular \wf need to be computed. The procedure will work if all 
such matrix elements can be shown to be finite.

\vskip .5 true cm
\noindent
{\bf Acknowledgements}

I thank B. Ananthanarayan for help with some of the numerics.

\vskip 1 true cm

\end{document}